\input harvmac
\input epsf
\newcount\figno
\figno=0
\def\fig#1#2#3{
\par\begingroup\parindent=0pt\leftskip=1cm\rightskip=1cm\parindent=0pt
\baselineskip=11pt
\global\advance\figno by 1
\midinsert
\epsfxsize=#3
\centerline{\epsfbox{#2}}
\vskip 12pt
{\bf Fig.\ \the\figno: } #1\par
\endinsert\endgroup\par
}
\def\figlabel#1{\xdef#1{\the\figno}}
\def\encadremath#1{\vbox{\hrule\hbox{\vrule\kern8pt\vbox{\kern8pt
\hbox{$\displaystyle #1$}\kern8pt}
\kern8pt\vrule}\hrule}}

\lref\rDb{J. Polchinski, "{\it Dirichlet Branes and Ramond-Ramond Charges,}" 
Phys. Rev. Lett. {\bf 75}, 4742 (1995) hep-th/9510017\semi
J. Dai, R.G. Leigh and J. Polchinski, "{\it New Connections between 
String Theories,}" Mod. Phys. Lett. {\bf A4}, 2073 (1989).}
\lref\rMyers{R.C. Myers, {\it Dielectric-Branes}, JHEP {\bf 9912}, 022 (1999), hep-th/9910053.}
\lref\rTsyetlin{A.A. Tseytlin, {\it Born-Infeld Action, Supersymmetry and String Theory,} in {\it The
Many Faces of the Superworld,} ed. M. Shifman (World Scientific, 2000), hep-th/9908105.}
\lref\rTseyt{A.A. Tseytlin, {\it On Non-Abelian generalization of Born-Infeld action in String Theory}, 
Nucl.Phys.{\bf B501}, 41 (1997), hep-th/9701125.}
\lref\rMyrsT{N.R. Constable, R.C. Myers and O. Tafjord, "{\it Noncommutative bion core}", Phys. Rev. {\bf D61}
106009 (2000) hep-th/9911136.}
\lref\MyersTh{N.R. Constable, R.C. Myers and O. Tafjord, "{\it Non-abelian brane intersections}", JHEP {\bf 0106:023}
(2001) hep-th/0102080.}
\lref\rGauntlett{J.P. Gauntlett, J. Gomis, and P.K. Townsend, "{\it BPS Bounds for Worldvolume Branes}", 
JHEP {\bf 9801} 003 (1998) hep-th/9711205\semi
D. Brecher, Phys. Lett. "{\it BPS States of the non-Abelian Born-Infeld Action}", {\bf B442} 117 (1998), hep-th/9804180.}
\lref\rRam{Pei-Ming Ho and Sanjaye Ramgoolam, "{\it Higher Dimensional Geometries from Matrix Brane 
Constructions,}" hep-th/0111278.}
\lref\rDoug{M.R. Douglas, "{\it Branes within Branes}," hep-th/9512077.}
\lref\rMba{D. Kabat and W. Taylor, "{\it Spherical Membranes in Matrix Theory,}," Adv.
Theor. Math. Phys. {\bf 2}
181 (1998), hep-th/9711078\semi 
J. Castelino, S. Lee and W. Taylor, "{\it Longitudinal Five-Branes as Four Spheres in Matrix Theory}," Nucl.
Phys. {\bf B256} 334 (1998), hep-th/9712105 .}
\lref\rep{J. Maldacena and A. Strominger, "{\it AdS3 Black Holes and a Stringy Exclusion Principle},"
{\bf JHEP 9812} 013 (2002), hep-th/9804085\semi
A. Jevicki and S. Ramgoolam, "{\it Noncommutative gravity from the AdS/CFT Correspondence}," {\bf JHEP 9904}
032 (1999), hep-th/9902059\semi
P.M. Ho and M. Li, "{\it Fuzzy Spheres in AdS/CFT Correspondence and Holography from Noncommutativity}," 
Nucl. Phys. {\bf B596} 259 (2001), hep-th/0004072\semi
M. Berkooz and H. Verlinde, "{\it Matrix Theory, AdS/CFT and Higgs-Coulomb Equivalence},"
{\bf JHEP 9912} 037 (1999), hep-th/9907100\semi
G. Bonelli, "{\it Matrix Strings in PP Wave Backgrounds from Deformed Super Yang-Mills
Theory}," JHEP {\bf 0208}:022 (2002), hep-th/0205213.}
\lref\rSR{S. Ramgoolam, "{\it On Spherical Harmonics for Fuzzy Spheres in Diverse Dimensions,}"
Nucl. Phys. {\bf B610} (2001) 461, hep-th/0105006\semi see also:
S. Ramgoolam, "{\it Higher dimensional geometries related to Fuzzy odd-dimensional spheres},"
hep-th/0207111.}
\lref\rYK{Y. Kimura, "{\it On Higher Dimensional Fuzzy Spherical Branes}," hep-th/0301055.}
\lref\rTA{T. Azuma and M. Bagnoud, "{\it Curved space classical solutions of a massive supermatrix model},"
hep-th/0209057.}


\Title{ \vbox {\baselineskip 12pt\hbox{}
\hbox{}  \hbox{June 2003}}}
{\vbox {\centerline{Non-Abelian BIonic Brane Intersections}
}}

\smallskip
\centerline{Paul L. H. Cook$^1$, Robert de Mello Koch$^{1,3}$ and 
Jeff Murugan$^{2,3}$}
\smallskip
\centerline{\it Department of Physics and Center for Theoretical Physics$^1$,}
\centerline{\it University of the Witwatersrand,}
\centerline{\it Wits, 2050, South Africa}
\centerline{\tt pcook, robert@neo.phys.wits.ac.za}
\smallskip
\centerline{\it Department of Mathematics and Applied Mathematics$^2$,}
\centerline{\it University of Cape Town,}
\centerline{\it South Africa}
\centerline{\tt jeff@hbar.mth.uct.ac.za}
\smallskip
\centerline{\it Stellenbosch Institute for Advanced Studies$^3$,}
\centerline{\it Stellenbosch,}
\centerline{\it South Africa}
\bigskip

{\vskip 20pt}

\noindent
We study "fuzzy funnel" solutions to the non-Abelian equations of motion of the D-string.
Our funnel describes ${n^6\over 360}$ coincident D-strings ending on ${n^3\over 6}$ D7-branes,
in terms of a fuzzy six-sphere which expands along the string. We also provide a dual description
of this configuration in terms of the world volume theory of the D7-branes. Our work makes use
of an interesting non-linear higher dimensional generalization of the instanton equations.


\Date{}

\def\MAKEdalamSIGN#1#2{%
\setbox0=\hbox{$\mathsurround 0pt #1{#2}$}
\dimen0=\ht0 \advance\dimen0 -0.8pt
\hbox{\vrule\vbox to\ht0{\hrule width\dimen0 \vfil\hrule}\vrule}}


\newsec{Introduction}

Many new results in string theory have been obtained by studying the low energy world volume theory of
D-branes\rDb. A fascinating example is the appearance of non-commutative geometry. In particular,
an interesting class of solutions has been obtained by studying a set of D1-branes that end on an
orthogonal D3-brane\rMyrsT\ or on an orthogonal D5-brane\MyersTh. These fuzzy funnel solutions consist 
of a fuzzy sphere geometry which expands along the length of the string. 

Fuzzy spheres themselves are a fascinating example of non-commutative geometry. They arise as 
solutions to matrix brane actions\rMba,\rMyrsT,\MyersTh\ and may also play a role in a spacetime explanation of the 
stringy exclusion principle\rep. The geometry of even dimensional fuzzy spheres has been investigated in 
\rRam\ and the detailed $SO(m)$ decomposition of the matrix algebras of the fuzzy spheres has been
given in \rSR.  For fuzzy spheres $S^m$ with $m>2$, it turns out that the matrix algebras contain more 
representations than is needed to describe functions on the sphere. In fact, in the classical limit
(limit of large matrices), the matrix algebras related to even dimensional fuzzy spheres approach
the algebra of functions of the higher dimensional space $SO(2k+1)/U(k)$. It has been argued that the 
appearance of these extra dimensions is a consequence of the Myers effect\rYK. 

In this article we study "fuzzy funnel" solutions to the non-Abelian equations of motion of the D-string.
Our funnel describes ${n^6\over 360}$ coincident D-strings ending on ${n^3\over 6}$ D7-branes. The geometry
of our solution is that of a fuzzy $S^6$ which expands along the string. This connection between the number 
of D-strings and the number of D7-branes has also been obtained directly from the non-commutative geometry of
the $S^6$. This solution is a natural generalization of the $D1\perp D3$\rMyrsT\ and 
the $D1\perp D5$\MyersTh\ solutions which
made use of the fuzzy $S^2$ and fuzzy $S^4$ respectively. We also provide a dual description of this 
configuration in terms of the world volume theory of the D7-branes. The D7-brane theory gauge field 
configurations have non-vanishing third Chern character on the six sphere surrounding the endpoints of the
D-strings. The energy, charge and radial profile of our solution computed in the two descriptions agree
exactly. 

Our paper is organized as follows: Since our solution makes use of the fuzzy $S^6$, we review the 
relevant matrix algebra in section 2. In section 3 we develop the description of our system using
the low energy D-string theory. In section 4 we recover the same results using the low energy D7-brane
theory. In section 5 we consider the simplest fluctuations on the fuzzy funnel solution.
Finally in section 6, we make some comments on the domains of validity of both the D-string and the
D7-brane theories.

\newsec{Fuzzy Six-Sphere}

In this section we review the construction of the fuzzy six-sphere. This is
done to establish notation and to derive a number of identities that will be
used in later sections. In preparing this section we found \rTA\ helpful.

To construct the fuzzy six-sphere, we need to construct solutions to the equation

\eqn\FSphere
{\sum_{i=1}^7 X^iX^i =c{\bf 1},}

\noindent
with $X^i$ a matrix, ${\bf 1}$ the identity matrix and $c$ a constant. Schur's lemma can be used
to obtain a simple construction of the matrices $X^i$. Toward this end, consider 
the Clifford algebra

$$\{\Gamma^i,\Gamma^j\}=2\delta^{ij},\qquad i,j=1,2,...,7.$$

\noindent
Denote the space on which the $\Gamma^i$ matrices act by $V$. The $n$-fold tensor 
product of $V$ is written as $V^{\otimes n}$. The $X^i$ are now obtained
by taking

$$ X^i =\left( \Gamma^i \otimes 1\otimes ...\otimes 1 +
1 \otimes \Gamma^i \otimes ...\otimes 1 +...+
1 \otimes 1\otimes ...\otimes \Gamma^i\right)_{st}.$$

\noindent
The subscript $st$ is to indicate that the above $X^i$ are to be restricted
to the completely symmetric and traceless tensor product space\foot{This restriction
is important if one is to obtain an irreducible representation, which is assumed 
in the application of Schur's lemma.}. To prove that the above $X^i$ do indeed provide
coordinates for the fuzzy six-sphere, one shows that $\sum_{i=1}^7 X^i X^i$ commutes
with the generators of $SO(7)$

$$ X^{kl}={1\over 2}\left(\big[\Gamma^k,\Gamma^l\big] \otimes 1\otimes ...\otimes 1 +
1 \otimes \big[\Gamma^k,\Gamma^l\big] \otimes ...\otimes 1 +...+
1 \otimes 1\otimes ...\otimes \big[\Gamma^k,\Gamma^l\big]\right)_{st}.$$

\noindent
The result \FSphere\ now follows from Schur's lemma. Further, using the Clifford 
algebra we easily find $c=n(n+6)$. The $X^{ij}$ matrices generate the $SO(7)$ Lie algebra. 
The matrix algebra associated with the fuzzy $S^6$ includes both the
$X^i$ and the $X^{ij}$. Together these matrices generate the $SO(7,1)$ Lie algebra. 

The symmetric traceless representation we work with has dimension

$$ N={1\over 360}(n+1)(n+2)(n+3)^2(n+4)(n+5), $$

\noindent
which identifies the representation generated by the $X^{ij}$ as the 
$\vec{r}=({n\over 2},{n\over 2},{n\over 2})$ irreducible representation of $SO(7)$.

Using the above definitions and the Clifford algebra, it is straightforward to derive the 
following identities (as usual, repeated indices are summed) 

$$\eqalign{\big[ X^{ij},X^{kl}\big]&=2\delta^{jk}X^{il}-2\delta^{ik}X^{jl}
+2\delta^{jl}X^{ki}-2\delta^{il}X^{kj}\cr
\big[X^{ij},X^k\big]&=2(\delta^{jk}X^i-\delta^{ik}X^j)\cr
X^iX^i&=c{\bf 1},\cr
X^{ij}X^j &=6X^i =X^jX^{ji}\cr
X^jX^{jk}X^{kl}X^l &=6^2 c{\bf 1}\cr
X^jX^{jk}X^{kl}X^{lm}X^{mn}X^n &=6^4 c{\bf 1}\cr
X^jX^{jk}X^{kl}X^{lm}X^{mn}X^{np}X^{pq}X^q &=6^6 c{\bf 1}\cr
X^{ij}X^{jl}&=6X^{il}-X^i X^l +c\delta^{il}{\bf 1},\cr
X^{ij}X^{ji}&=6c{\bf 1}\cr
X^{ij}X^{jk}X^{kl}X^{li}&=6c^2{\bf 1}\cr
X^{ij}X^{jk}X^{kl}X^{lm}X^{mn}X^{ni}&=6c^3{\bf 1}\cr
\epsilon^{ijklmnq}X^i X^j X^k X^l X^m X^n &=i(384+288n+48n^2)X^q.}$$

The geometry of the fuzzy six sphere has been studied in detail in \rRam. These authors
argue that the fuzzy six sphere is a bundle over the sphere $S^6$. In the classical limit
the fibre over the sphere is the symmetric space $SO(6)/U(3)$. The result of relevance to
us, following from this geometrical analysis, is that one can identify points in the base,
and as a consequence it is possible to read off the 6-brane charge as ${1\over 6}(n+1)(n+2)(n+3)$.
We will see that it is possible to reproduce this purely non-commutative geometric derivation of 
the charge using either a dynamical analysis based on the non-Abelian Born-Infeld description of
$N$ coincident D-strings in the large $N$ limit, or by using the non-Abelian Born-Infeld description 
of $n^3/6$ D7-branes, in the large $n$ limit. The D6 brane charge will correspond to a 
D7-brane charge in our T-dual description.

In the remainder of this article we work in the large $n$ limit. Consequently we use

$$ N={n^6\over 360},\qquad c=n^2$$

\noindent
and take the 6-brane charge to be $n^3/6$. 

\newsec{Description of the D1 $\perp$ D7 system in terms of $N$ D1 branes}

In this section we study the fuzzy geometry of the $D7\perp D1$ system, using 
the non-Abelian theory describing $N$ coincident D-strings. Our construction 
employs the fuzzy six-sphere to construct a fuzzy funnel in which the D-strings 
expand into orthogonal D7-branes. We use an approach based on minimizing the 
energy\rGauntlett, which generalizes the results obtained in \rMyrsT\ for D1-branes
expanding into orthogonal D3-branes and the results in \MyersTh\ for D1-branes
expanding into orthogonal D5-branes.

The low energy effective action for $N$ D-strings is given by the non-Abelian 
Born-Infeld action\rMyers,\rTsyetlin

\eqn\Nabi
{S=-T_1\int d^2\sigma STr\sqrt{-\det\left[
\matrix{\eta_{ab} &\lambda\partial_a\Phi^j\cr 
-\lambda\partial_b\Phi^i &Q^{ij}}
\right]}\equiv -T_1\int d^2\sigma STr\sqrt{-\det M}\quad ,}

\noindent
where

$$ Q^{ij} = \delta^{ij}+i\lambda\big[\Phi^i ,\Phi^j\big],\qquad
\lambda =2\pi l_s^2.$$

\noindent
The symmetrized trace prescription \rTseyt\ (indicated by $STr$ in the above action)
instructs us to symmetrize over all
permutations of $\partial_a\Phi^i$ and $\big[\Phi^i ,\Phi^j\big]$ within the trace over
the gauge group indices, after expanding the square root. We are using static gauge so
that the worldsheet coordinates are identified with spacetime coordinates as
$\tau =x^0$ and $\sigma =x^9$. The transverse coordinates are now the non-Abelian
scalars $\Phi^i$, $i=1,...,8$. These scalars are $N\times N$ matrices transforming in the 
adjoint representation of the $U(N)$ gauge symmetry present on the worldsheet of the D1s. 

We seek static solutions with seven of the scalars excited. It is a tedious but
straightforward exercise to show that it is consistent with the equations of 
motion to make use of a static ansatz that involves seven of the scalars, at the 
level of the action. With this ansatz and a rather lengthly calculation we obtain

$$\eqalign{-\det (M)&=1+{\lambda^2\over 2}\Phi^{ij}\Phi^{ji}
+{\lambda^4\over 8}(\Phi^{ij}\Phi^{ji})^2-{\lambda^4\over 4}
\Phi^{ij}\Phi^{jk}\Phi^{kl}\Phi^{li}\cr
&\,\,\, +\lambda^6\left({(\Phi^{ij}\Phi^{ji})^3\over 48}\right.
-{\Phi^{mn}\Phi^{nm}\Phi^{ij}\Phi^{jk}\Phi^{kl}\Phi^{li}\over 8}
+\left.{\Phi^{ij}\Phi^{jk}\Phi^{kl}\Phi^{lm}\Phi^{mn}\Phi^{ni}\over 6}\right)\cr
&\,\,\, +\lambda^2\partial_\sigma\Phi^i\partial_\sigma\Phi^i+
\lambda^4\left({\partial_\sigma\Phi^k\partial_\sigma\Phi^k
\Phi^{ij}\Phi^{ji}\over 2}-\partial_\sigma\Phi^i\Phi^{ij}\Phi^{jk}
\partial_\sigma\Phi^k\right)\cr
&\,\,\, -\lambda^6\left(
{\partial_\sigma\Phi^m\partial_\sigma\Phi^m
\Phi^{ij}\Phi^{jk}\Phi^{kl}\Phi^{li}\over 4}
-{\partial_\sigma\Phi^i\partial_\sigma\Phi^i
(\Phi^{ij}\Phi^{ji})^2
\over 8}\right.\cr
&\,\,\, +{\partial_\sigma\Phi^i
\Phi^{ij}\Phi^{jk}\partial_\sigma\Phi^k
\Phi^{ml}\Phi^{lm}\over 2} -\partial_\sigma\Phi^i
\Phi^{ij}\Phi^{jk}\Phi^{kl}\Phi^{lm}
\partial_\sigma\Phi^m\Big)\cr
&\,\,\, -\lambda^8\left(
-{\partial_\sigma\Phi^k\partial_\sigma\Phi^k
(\Phi^{ij}\Phi^{ji})^3\over 48}\right.+{\partial_\sigma\Phi^p\partial_\sigma\Phi^p
\Phi^{ij}\Phi^{ji}\Phi^{kl}\Phi^{lm}\Phi^{mn}\Phi^{nk}\over 8}\cr
&\,\,\, -{\partial_\sigma\Phi^p\partial_\sigma\Phi^p
\Phi^{ij}\Phi^{jk}\Phi^{kl}\Phi^{lm}\Phi^{mn}\Phi^{ni}\over 6}
+{\partial_\sigma\Phi^i\Phi^{ij}\Phi^{jk}\partial_\sigma\Phi^k
(\Phi^{ml}\Phi^{lm})^2\over 8}\cr
&\,\,\, -{\partial_\sigma\Phi^i\Phi^{ij}\Phi^{jk}\partial_\sigma\Phi^k
\Phi^{ml}\Phi^{ln}\Phi^{np}\Phi^{pm}\over 4}-{\partial_\sigma\Phi^i
\Phi^{ij}\Phi^{jk}\Phi^{kl}\Phi^{ln}\partial_\sigma\Phi^n
\Phi^{mp}\Phi^{pm}\over 2}\cr
&\,\,\, +\partial_\sigma\Phi^i
\Phi^{ij}\Phi^{jk}\Phi^{kl}\Phi^{ln}\Phi^{np}\Phi^{pm}
\partial_\sigma\Phi^m\Big),}$$

\noindent
where

$$ \Phi^{ij}=\big[\Phi^i,\Phi^j\big].$$

\noindent
Our ansatz for the funnel solution is given by

$$\Phi^i=R(\sigma )X^i .$$

\noindent
We have checked that the equation determining $R(\sigma )$ obtained by substituting this ansatz into the
equations of motion (following from \Nabi) agree with the equations obtained by inserting this ansatz into
the action \Nabi\ and varying with respect to $R(\sigma )$. Following this second procedure, inserting
the above ansatz into \Nabi\ we obtain

\eqn\Acttwo
{\eqalign{S&=-T_1\int d^2\sigma STr\sqrt{\left(1+\left({d\bar{R}\over d\sigma}\right)^2\right)
\left(1+ f(\bar{R}) \right)},\cr
f(\bar{R})&=12{\bar{R}^4\over c\lambda^2}
+48{\bar{R}^8\over c^2\lambda^4}+
64{\bar{R}^{12}\over c^3\lambda^6},}}

\noindent 
where we have introduced the physical radius

$$\bar{R}=\sqrt{c}\lambda R .$$

\noindent
In obtaining this result, use has been made of the identities listed in section 2. The formula
\Acttwo\ is not exact - it catches only the leading large $N$ contribution. If we expand the square
root in \Nabi\ and implement the symmetrization of the trace for each term in the expansion, we
find corrections to \Acttwo\ of order $1/c$ relative to the leading term. Thus, our results are only
valid for large $N$.

Since this is a static configuration, it is easy to obtain the following expression for the energy
of our solution

$$\eqalign{ E &=NT_1\int d\sigma
\sqrt{\left(1+\left({d\bar{R}\over d\sigma}\right)^2\right)\left(1+f(\bar{R})\right)}\cr
&=NT_1\int d\sigma\sqrt{\left({d\bar{R}\over d\sigma}\pm
\sqrt{f(\bar{R})}\right)^2 + \left(1\mp{d\bar{R}\over d\sigma}
\sqrt{f(\bar{R})}\right)^2}\cr
&\ge NT_1\int d\sigma \left(1\mp{d\bar{R}\over d\sigma}\sqrt{f(\bar{R})}
\right).}$$

\noindent
The above inequality is saturated when

$$0={d\bar{R}\over d\sigma}\pm\sqrt{{12\bar{R}^4\over c\lambda^2}
+{48\bar{R}^8\over c^2\lambda^4}+{64\bar{R}^{12}\over c^3\lambda^6}}.$$

\noindent
For small $\bar{R}$ it is simple to obtain

$${d\bar{R}\over d\sigma}=\mp{2\sqrt{3}\bar{R}^2\over\sqrt{c}\lambda}\qquad
\Rightarrow\qquad
\bar{R}=\pm{\sqrt{c\lambda}\over 2\sqrt{3} (\sigma-\sigma_0)}.$$

\noindent
This is the same behaviour as was found in both the D3-brane funnel\rMyrsT, and the D5-brane
funnel\MyersTh. We have reproduced the expected behaviour for any D-string funnel in the region 
where the funnel is well approximated by the D-string. Consider now the large $\bar{R}$ region.
If our funnel is to expand into an orthogonal D7-brane at large $\bar{R}$, the expansion must
be given by an harmonic function in seven spatial dimensions. At large $\bar{R}$ we find

$${d\bar{R}\over d\sigma}=\mp{8\bar{R}^6\over c^{3\over 2}\lambda^3}\qquad
\Rightarrow\qquad
\sigma-\sigma_0=\pm{c^{3/2}\lambda^3\over 40\bar{R}^5},$$

\noindent
which is indeed the correct harmonic behaviour needed for a D7-brane to appear at $\sigma=\sigma_0$.

Further evidence that we have a funnel expanding into coincident D7-branes is provided by computing
the RR charge and energy of this solution. The energy of our solution is

$$\eqalign{E&= NT_1\int d\sigma \left(1 +{d\bar{R}\over d\sigma}\sqrt{{12\bar{R}^4\over c\lambda^2}
+{48\bar{R}^8\over c^2\lambda^4}+{64\bar{R}^{12}\over c^3\lambda^6}}
\right)\cr
&=NT_1\int_0^\infty d\sigma
+NT_1\int_0^\infty d\bar{R}\sqrt{{12\bar{R}^4\over c\lambda^2}
+{48\bar{R}^8\over c^2\lambda^4}+{64\bar{R}^{12}\over c^3\lambda^6}}.}$$

\noindent
The first term is easily identified as the energy of $N$ semi-infinite D-strings stretching from
$\sigma=0$ to $\sigma=\infty$. Now consider the second term. We compute this term for large $\bar{R}$,
where we expect that the funnel is expanding into a number of coincident D7-branes. Using the identities

$$ N={n^6\over 360},\qquad c=n^2,$$

\noindent
which are valid for large $n$, as well as the known relation between the tension of the D-string and the
D7-brane and of the D-string and the D3-brane

$$ T_7={T_1\over (2\pi l_s)^6},\qquad T_3={T_1\over (2\pi l_s)^2},$$

\noindent
it is straightforward to obtain the following result for the energy

\eqn\DOenergy
{E=NT_1\int_0^\infty d\sigma +{n^3\over 6}T_7\Big({16\pi^3 \over 15}\int d\bar{R}\,\,\bar{R}^6\Big)
+{n^5\over 240}T_3\int d\bar{R}\,\, 4\pi\bar{R}^2 +\Delta E,}

\noindent
where

$$\Delta E=NT_1 c^{1\over 4}\sqrt{\lambda\over 2}\int_0^\infty\left[
\sqrt{u^{12} +3u^8 +3u^4}-u^6-{3\over 2}u^2\right]du \approx (0.2629...)NT_1
c^{1\over 4}\sqrt{\lambda}.$$

\noindent
The second term in \DOenergy\ is precisely the energy of ${n^3\over 6}$ D7-branes, so that we have reproduced
the non-commutative geometric derivation of the charge given in \rRam. The two terms given provide the analog
of the two terms providing the total energy of the supersymmetric $D3\perp D1$ system\rMyrsT. The fact that there are 
further contributions to the energy matches what one finds in the analysis of the $D5\perp D1$ system\MyersTh. In the
$D5\perp D1$ context, this was interpreted as a consequence of the fact that the system is not supersymmetric. 
The third term in \DOenergy\ is apparently the energy of ${n^5\over 240}$ D3 branes. Recall that the zero scale 
size limit of an instanton in a Dp-brane corresponds to a D(p-4) brane bound to the Dp-brane\rDoug. Thus, this 
term is naturally interpreted as an instanton contribution in the D7-brane theory. It is interesting to note that 
the corresponding term in the $D5\perp D1$ system arises from a D1 contribution, which can be interpreted as an 
instanton contribution in the D5-brane theory.  It would be interesting to understand the physical origin of 
this term, perhaps as a consequence of the Myers effect. The last term represents a finite binding energy.

We have evidence that our solution describes a funnel expanding into a number of coincident D7-branes located
at $\sigma=0$. The D7 branes expand to fill the $X^i$, $i=1,2,...,7$ directions. If this is indeed the case, this 
configuration should be a source for the eight-form RR-potential $C^{(8)}_{012345678}$. We check this, providing
a further check of the D7-brane charge computed by studying the energy of our configuration. The relevant source 
term comes from the following contribution to the non-Abelian Wess-Zumino action 

$$ S_{WZ}=-i{\lambda^3\over 6}\mu_1\int \
STr\,\, P\big[ (\dot{1}_\Phi \dot{1}_\Phi )^3C^{(8)}\big] .$$

\noindent
Evaluating the value of this term for our solution

$$\eqalign{
S_{WZ}&=-i{\lambda^4\over 6}\mu_1 \int d\sigma d\tau
C^{(8)}_{01234567}\,\, STr \big(\epsilon^{ijklmnp}\Phi^i
\Phi^j\Phi^k\Phi^l\Phi^m\Phi^n\partial_\sigma\Phi^p\big)\cr
&=-i{\lambda^4\mu_1\over 6\lambda^7 c^{7/2}}\int d\sigma d\tau
C^{(8)}_{01234567}\,\, STr \big(\epsilon^{ijklmnp}G^i
G^j G^k G^l G^m G^n G^p\big)\bar{R}^6{d\bar{R}\over d\sigma},}$$

\noindent
using the identities given in section 2, the relation between D7 and D1 charges

$$\mu_7={\mu_1\over (2\pi l_s)^6},$$

\noindent
and working in the large $n$ limit, we obtain

$$ S_{WZ} ={n^3\over 6}\mu_7\left({16\pi^3\over 15}\int d\bar{R}\,\,
C^{(8)}_{01234567}\,\,\bar{R}^6 \right) .$$

\noindent
This is exactly the seven-brane source term we would expect to get, if we have $n^3/6$
D7-branes, in complete agreement with our energy computation. 

Up to now, we have obtained solutions by employing a method which minimizes the energy. We end
this section with a direct analysis of the equations of motion. Requiring that \Acttwo\ is stationary
with respect to variations of $\bar{R}$, we obtain the following equation of motion

$$ \sqrt{1+\left({d\bar{R}\over d\sigma}\right)^2}{d\sqrt{1+f(\bar{R})}\over d\bar{R}}=
{d\over d\sigma}\left(
{\sqrt{1+f(\bar{R})\over 1+({d\bar{R}\over d\sigma})^2}{d\bar{R}\over d\sigma}}
\right).$$

\noindent
After some straightforward manipulations, this equation of motion can be written as

$$\left({d\bar{R}\over d\sigma}\right)^{-1}{d\over d\sigma}\left(
\sqrt{1+f(\bar{R})\over 1+({d\bar{R}\over d\sigma})^2}
\right)=0,$$

\noindent
which is easily integrated to give

\eqn\GenProf
{{d\bar{R}\over d\sigma}=\pm\sqrt{kf(\bar{R})-1},}

\noindent
where $k$ is a non negative dimensionless constant of integration. For $k=1$, we reproduce the energy we 
obtained above by minimizing the energy. For $0\le k\le 1$, the solution reaches $\bar{R}=0$ at finite
value of $\sigma$ so that the funnel "pinches" off. As explained in \rMyrsT\ this solution can naturally
be continued past $\bar{R}=0$, by matching to a second pinched off funnel. This configuration provides
the description of two parallel sets of coincident D7-branes, joined by $N$ finite length D-strings. If $k>1$, the 
solution reaches ${d\bar{R}\over d\sigma}=0$ at finite $\sigma$ and terminates. Again \rMyrsT, this solution
is naturally continued by matching to a second funnel. In this case, the double funnel describes $N$ finite
D-strings joining a set of coincident anti-D7 branes with a set of parallel coincident D7-branes.

This concludes our discussion of the D-string theory. In the next section, we turn to a dual description
of the same configuration, which employs the non-Abelian world volume theory of the coincident D7-branes.

\newsec{D1 $\perp$ D7 configuration using a D7 world volume description}

In the previous section we have argued that our funnel describes $N={n^6\over 360}$ D-strings expanding into
${n^3\over 6}$ D7-branes. Consequently the D7-brane world volume theory is a $7+1$ dimensional non-Abelian Born-Infeld
theory with gauge group $U({n^3\over 6})$. Further, to describe the D-strings, we will also have to excite one of the 
transverse scalars. This scalar has to reside in the overall $U(1)$ component of the $U({n^3\over 6})$ gauge group,
since it describes a deformation of the geometry of all of the D7-branes. Consequently, we consider the action

$$S=-T_7\int d^8\sigma STr\sqrt{-\det (G_{ab}+\lambda^2\partial_a\phi\partial_b\phi
+\lambda F_{ab})}.$$

\noindent
We employ spherical coordinates on the D7 worldvolume

$$ ds^2=G_{ab}d\sigma^a d\sigma^b =-dt^2 +dr^2 +r^2 g_{ij}d\alpha^i d\alpha^j ,$$

\noindent
with $g_{ij}$ the metric on the six sphere of unit radius, $r$ is the radial coordinate and $\alpha^i$ the angles.
In analogy to the $D5\perp D1$ system\MyersTh, we make the following ansatz for the scalar and gauge fields

$$\phi =\phi(r),\qquad A_r=0,\qquad A_{\alpha^i}=A_{\alpha^i}(\alpha^j).$$

\noindent
Once again we have examined the full equations of motion and have verified that this is indeed a consistent
ansatz. Inserting this ansatz into the above action, we obtain

\eqn\DFaction
{\eqalign{S_7&=-T_7\int d^8\sigma \sqrt{\left(1+\lambda^2 
\left({d\phi\over dr}\right)^2\right)g}STr\sqrt{h(r)},\qquad
g=\det g_{ij}\cr
&=-T_7\int d^8\sigma L_7\cr
h(r)&=r^{12}+{1\over 2}r^8\lambda^2F^{ij}F_{ij}+{1\over 128}r^4\lambda^4
\epsilon_{ijklmn}\epsilon^{ijopqr}F^{kl}F^{mn}F_{op}F_{qr}\cr
&\,\,\, +{1\over 2304}\lambda^6(\epsilon_{ijklmn}F^{ij}F^{kl}F^{mn})^2 .}}

\noindent
In the above expression, $F_{ij}\equiv F_{\alpha^i\alpha^j}$, indices on the field strength are raised and lowered 
with the metric $g_{ij}$, and $\epsilon_{123456}=g$. The equation of motion for the scalar is

$${d\over dr}\left({\partial L_7\over\partial\phi'}\right)=0, \qquad \phi' ={d\phi\over dr}.$$

\noindent
This is easily integrated to obtain

$${\lambda^2 \phi'\over\sqrt{1+\lambda^2 (\partial_r\phi)^2}}=
{f(\alpha^i)\over\sqrt{g}STr\sqrt{h(r)}},$$

\noindent
where $f(\alpha^i)$ is an arbitrary function of integration depending only on the angles $\alpha^i$.
The left hand side of the above equation is independent of the $\alpha^i$, so we must have
$STr\sqrt{h(r)}$ independent of the angles and further,

$$ f(\alpha^i )={\sqrt{g}\lambda^4\over b},$$

\noindent
with $b$ a dimensionless constant. With this choice we obtain

\eqn\DFprof
{\lambda\phi' =\pm{1\over\sqrt{{b^2 \left(STr \left(\sqrt{h(r)}\right)\right)^2\over\lambda^6}-1}}.}

\noindent
After identifying $\sigma=\lambda\phi$ we have 

$$\lambda {d\phi\over dr}={d\sigma\over dr}.$$

\noindent
With this identification and $r=\bar{R}$, the radial profile \DFprof\ can be matched to the result we obtained from 
the D-string world volume theory \GenProf\ by setting

$$ k f(r) ={\left(STr \sqrt{h(r)}\right)^2 b^2\over \lambda^6}.$$

\noindent
This last condition can be satisfied by choosing

\eqn\relnn
{\eqalign{F^{ij}F_{ij}&={3c\over 2}{\bf 1},\cr
\epsilon_{ijklmn}\epsilon^{ijopqr}F^{kl}F^{mn}F_{op}F_{qr}&=24 c^2{\bf 1},\cr
(\epsilon_{ijklmn}F^{ij}F^{kl}F^{mn})^2&=36 c^3{\bf 1},}}

\noindent
where ${\bf 1}$ is the ${n^3\over 6}\times {n^3\over 6}$ unit matrix. It is interesting to note that
these last three identities reduce to a single independent equation if one chooses

\eqn\reln
{8\sqrt{c}F^{ij}=\epsilon^{ijklmn}F_{kl}F_{mn} .}

\noindent
This last equation provides an interesting non-linear higher dimensional generalization of the
instanton equation.
This relation is also suggested by the D-string description\rRam. In Matrix Theory, the commutator 
$X^{\mu\nu}=i\big[X^\mu,X^\nu\big]$ of the matrix valued coordinates is naturally interpreted 
as a field strength. The state for which

$$ X^7|s\rangle =-n|s\rangle, \qquad X^i|s\rangle =0,\quad i<7,$$

\noindent
corresponds to a point at the north pole of the sphere. Locally at the north pole, directions $i$
with $i<7$ correspond to the $\alpha^i$ directions. Acting on this state, we find that the only non-zero 
"field strengths" are

$$ i\big[X^1,X^2\big]|s\rangle =2n|s\rangle,\qquad
i\big[X^3,X^4\big]|s\rangle =-2n|s\rangle,\qquad
i\big[X^5,X^6\big]|s\rangle =-2n|s\rangle .$$

\noindent
Since $\sqrt{c}=n$, we see that the field strengths at the north pole do indeed satisfy \reln. In the remainder
of this section we will assume that our field strengths satisfy \relnn\ and \reln. We will not address the issue 
of obtaining an actual gauge field solution from which we can compute these field strengths. 
Note that the above field strengths satisfy

$$ {1\over 48\pi^3}\int Tr\left({\epsilon_{ijklmn}F^{ij}F^{kl}F^{mn}\over 8}\right)\sqrt{g}d^6\alpha
={n^6\over 360}=N,$$

\noindent
exactly as one would expect for any six sphere surrounding the D-string endpoints.

We now turn to a computation of the energy of this solution. To compare to the energy of the configuration that 
saturated the energy bound, we now set $k=1$. The energy is

$$E=T_7\int\sqrt{g}d^6\alpha dr\sqrt{1+\lambda^2\left( {d\phi\over dr}\right)^2}{n^3\over 6}
\sqrt{r^{12} +{3r^8\lambda^2 c\over 4}+{3r^4\lambda^4 c^2\over 16}+{\lambda^6 c^3\over 64}}.$$

\noindent
After using \DFprof\ this becomes

$$\eqalign{
E&=T_1{n^6\over 360}\int_0^\infty dr\lambda{d\phi\over dr}+T_7{16\pi^3\over 15}{n^3\over 6}
\int_0^\infty dr\sqrt{r^{12} +{3r^8\lambda^2 c\over 4}+{3r^4\lambda^4 c^2\over 16}}\cr
&=NT_1\int_0^\infty d\sigma +{n^3\over 6}T_7\Big({16\pi^3 \over 15}\int d\bar{R}\,\,\bar{R}^6\Big)
+{n^5\over 240}T_3\int d\bar{R}\,\, 4\pi\bar{R}^2 +\Delta E,}$$

\noindent
where again

$$\Delta E=NT_1 c^{1\over 4}\sqrt{\lambda\over 2}\int_0^\infty\left[
\sqrt{u^{12} +3u^8 +3u^4}-u^6-{3\over 2}u^2\right]du \approx (0.2629...)NT_1
c^{1\over 4}\sqrt{\lambda}.$$

\noindent
This exactly matches the energy computed using the D-string description. Thus, the energy, radial profile of the
funnel and charge computed using the D7 world volume theory is in exact agreement with the calculations performed
using the D-string world volume theory.

\newsec{Fluctuations}

In this section we study the propagation of fluctuations on the fuzzy funnel solution obtained in section 3.
For a similar analysis of fluctuations for the $D3\perp D1$  and the $D5\perp D1$ systems see \rMyrsT\ and 
\MyersTh\ respectively.

Since our funnel has the topology $R\times S^6$, the fluctuations of this geometry are naturally decomposed
in terms of the spherical harmonics on the $S^6$. Of course, we have a fuzzy $S^6$, so it is natural to expand
the fluctuations in terms of traceless symmetric products of the $X^i$, which provide the deformation of the
usual algebra of functions on $S^6$. One consequence of the fact that we use a fuzzy sphere is simply that there
is a highest angular momentum $l\le l_{max}=n$. Concretely, we consider fluctuations of the form

$$ \delta \Phi^8 ={\cal C}_{i_1 i_2 ...i_n}(\tau,\sigma)X^{i_1}X^{i_2}...X^{i_n},\qquad
\delta\Phi^i=0,\quad i<8.$$

\noindent
where ${\cal C}_{i_1 i_2 ...i_n}(\tau,\sigma)$ is required to be a traceless symmetric tensor.
Our goal in this section is simply to show that these modes, which correspond to partial waves of angular momentum
$n$, see the correct angular momentum barrier. 

The lowest order equation of motion is

$$(-\partial_\tau^2+\partial_\sigma^2)\Phi^i=\big[\Phi^j,\big[\Phi^j,\Phi^i\big]\big].$$

\noindent
This equation of motion is valid for small $\Phi^j$ and hence corresponds to the region of small
$R(\sigma )$. The linearized equation for the fluctuation following from this lowest order equation 
of motion is

$$(-\partial_\tau^2+\partial_\sigma^2)\delta\Phi^8=\big[\delta\Phi^j,\big[\Phi^j,\Phi^8\big]\big]
+\big[\Phi^j,\big[\delta\Phi^j,\Phi^8\big]\big]
+\big[\Phi^j,\big[\Phi^j,\delta\Phi^8\big]\big].$$

\noindent
Since $\Phi^8=0$ and $\delta\Phi^j=0$ for $j<8$, this simplifies to

\eqn\fluct
{(-\partial_\tau^2+\partial_\sigma^2)\delta\Phi^8=
\big[\Phi^j,\big[\Phi^j,\delta\Phi^8\big]\big].}

\noindent
To evaluate the right hand side, we need to use the result

$$\eqalign{\big[\Phi^j,\big[\Phi^j,\delta\Phi^8\big]\big]&=R^2(\sigma)C^{i_1 i_2 ...i_n}\big[G^j,\big[
G^j,G^{i_1}G^{i_2}...G^{i_n}\big]\big]\cr
&=4n(n+4)R^2(\sigma)C^{i_1 i_2 ...i_n}G^{i_1}G^{i_2}...G^{i_n}.}$$ 

\noindent
Identifying $R^2(\sigma)={1\over 12\sigma^2}$ which is valid when $R(\sigma )$ is small, we obtain

$$\left(\partial_\tau^2-\partial_\sigma^2+{n(n+4)\over 3\sigma^2}\right)C^{i_1 i_2 ...i_n}(\tau,\sigma)=0.$$

\noindent
Thus, the double commutator on the right hand side of \fluct\ has indeed reproduced the correct angular
momentum barrier.

\newsec{Final Comments} 

We have obtained a description of the $D1\perp D7$ system in terms of a fuzzy six-sphere which 
expands along the string. We have studied the energy, charge and radial profile of this configuration
using the non-Abelian equations of motion of the D-string and also by using the dual description provided
by the world volume theory of the D7-branes. Our analysis is limited to the low energy world volume
theory in each case. The agreement between descriptions is perfect.
Further, we have found that the configuration describes ${n^6\over 360}$ coincident D-strings ending on 
${n^3\over 6}$ D7-branes. This relation between the number of D-strings and D7-branes has also been
obtained from a direct study of the non-commutative geometry of the fuzzy $S^6$.

This precise agreement between the two descriptions is also a feature of the $D1\perp D3$ and
$D1\perp D5$ systems. For the system we have studied in this article, we'd expect the D7-brane
world volume theory will provide a reliable description for those regions of the funnel that 
have opened up to fill out a seven dimensional spatial volume and are hence well approximated as
a D7-brane. The D-string world volume theory should provide a reliable description of the funnel
in the regions where the funnel is very thin and hence well approximated by a D-string. Thus we 
have two complementary descriptions of the $D1\perp D7$ system. How are we to understand the 
agreement between the two descriptions of the $D1\perp D7$ system? 

There are two potential sources of corrections to both descriptions. There are both higher derivative
corrections and higher order commutator corrections. Following \MyersTh\ we assume that we can ignore
higher derivative corrections when $l_s|\partial^2\Phi|<\!\! <|\partial\Phi|$. For the D-string theory, 
we easily find that this condition implies that $r<\!\! < \left({n^3\pi^3\over 12}\right)^{1\over 5}l_s$.
For the D7-brane theory this condition implies that $r>\!\! >2l_s$. Thus, for large $n$ there is a significant
region ($2l_s<\!\! <r<\! \!<\left({n^3\pi^3\over 12}\right)^{1\over 5}l_s$) where both descriptions do not
receive higher derivative ($\alpha'$) corrections. A conservative bound for the region in which higher commutator
terms are avoided is obtained by requiring that the Taylor expansion of the square root in the D-string
action should converge very rapidly. This implies that $r<\!\! < \sqrt{n}l_s$. For large $n$ we have
$\sqrt{n}>\!\! > \left({n^3\pi^3\over 12}\right)^{1\over 5},$ so that this is less restrictive than what we 
obtained above. We have not established the analogous region in which higher commutator
terms are avoided in the D7-brane theory.

$$ $$

{\it Acknowledgements:} We would like to thank Sanjaye Ramgoolam for extremely helpful 
discussions and for comments on the manuscript. 
We would also like to thank Rocco Duvenhage, Phil Ferrer 
and Joao Rodrigues for pleasant discussions.
The work of RdMK and PC is supported by NRF grant number Gun 2047219. The work of JM is supported by 
the Lindbury trust and a research associateship of the University of Cape Town. RdMK and JM would like 
to thank Fritz Hahne, Bernard Lategan and Maria Mouton
for their warm hospitatility at the Stellenbosch Institute 
for Adavanced Studies, where this work was completed.

\listrefs
\vfill\eject
\bye